\documentclass[conference]{IEEEtran}

\usepackage{cite}
\usepackage{amsmath,amssymb,amsfonts}
\usepackage{algorithmic}
\usepackage{graphicx}
\usepackage{textcomp}
\usepackage{xcolor}
\def\BibTeX{{\rm B\kern-.05em{\sc i\kern-.025em b}\kern-.08em
    T\kern-.1667em\lower.7ex\hbox{E}\kern-.125emX}}
\usepackage{tikz}
\usetikzlibrary{decorations.pathreplacing, arrows.meta}

\begin{document}

\title{Beyond the IT Checklist: Engineering a Reasonable Standard of Care for Cyber Safety}

\author{
\IEEEauthorblockN{Matthew E. Jablonski, Linton Wells II, 
Kathryn B. Laskey, F. Brett Berlin}
\IEEEauthorblockA{\textit{College of Engineering and Computing} \\
\textit{George Mason University} \\
Fairfax, VA \\
\textit{\{mjablons, lwells, klaskey, fberlin\}@gmu.edu}}
}

\maketitle

\begin{abstract}
Current U.S. cyber policy, centered on security, often treats documentation of controls and incident reports as a proxy for safety in the built environment. This paper argues that such an approach is inadequate for cyber-physical systems, where digital failures can produce kinetic harm. We construct and code a corpus of critical infrastructure policy documents ($N=292$, 2000--2025) to examine how ``reasonable care'' is operationalized across the NIST SP 800-160 Vol.~2 resilience lifecycle. The resulting maps show that obligations are concentrated in the \textit{Anticipate} phase and emphasize administrative compliance, while \textit{Withstand} and \textit{Recover} phases rely heavily on delegated references to IT-focused control catalogs that are poorly aligned with physics-based hazards. We identify three major disconnects: miscalibrated delegated standards, recovery defined as notification rather than engineered navigation, and uneven adaptation requirements across sectors. We then propose a modernized standard of care anchored in hazard-specific traceability, structured assurance cases, and cyber resiliency engineering. Finally, we recommend that federal policy pair these engineering obligations with targeted incentives so that resilient architectures for critical infrastructure become a viable business decision rather than an unfunded expectation.
\end{abstract}


\section{Introduction}
\label{sec:introduction}

Operational Technology (OT) and Information Technology (IT) have converged to turn the built environment into a cyber-physical contested environment with conflicting objectives.
Current U.S. policy ignores this hybrid reality. 
Rather than treating Cyber Safety as a distinct engineering discipline, regulators overlay data-centric IT security controls onto physics-centric systems. 
This creates a structural gap where asset owners can be fully compliant with major federal standards even though their systems are not engineered to tolerate failures or attacks.

The failure of this compliance-centric model is empirically evident in the defense industrial base. 
A 2025 industry survey found that while 69\% of contractors claimed National Institute of Standards and Technology (NIST) Special Publication (SP) 800-171 compliance, only 30\% passed verified assessments~\cite{CyberSheath2025}. 
Concurrently, malware infections at major defense contractors~\cite{HudsonRock2025} demonstrate that IT-focused checklists do not ensure the engineering rigor required to prevent adversary access.

This failure stems from a misapplication of ``reasonable care.''
Legally defined as the prudence of a rational person~\cite{LII_Reasonable_Care}, this standard is typically interpreted in cybersecurity as adherence to accepted best practices~\cite{CIS_Reasonable_Cyber_2024}, such as NIST Cybersecurity Framework (CSF) or Cybersecurity Maturity Model Certification (CMMC).
However, for Cyber Safety, where digital failure creates physical harm, administrative adherence is insufficient.
Therefore, we argue that in the context of cyber-physical systems, \textit{Reasonable Care} requires the verifiable application of systems security engineering (SSE) principles---specifically \textit{traceability}, \textit{assurance}, and \textit{resilience}---to ensure that a system remains safe even when it is not secure.

We argue that a valid standard of care for the built environment must shift from administrative compliance checklists in favor of systems engineering rigor.
Through a policy mapping of the U.S. critical infrastructure policy corpus (2000–2025), we classify requirements by obligation (Administrative, Coordination, or Engineering). 
The analysis reveals a systemic ``Verification Gap'': while policies mandate activity (plans, meetings, notifications), they lack mandates for efficacy (engineering evidence that the hazard is neutralized).

We propose closing this gap with a three-part engineering standard, (1) traceability from hazard to control (e.g., Cyber-Informed Engineering~\cite{Bochman2021}), (2) structured assurance cases (e.g., ISO/IEC/IEEE 15026-2~\cite{ISO15026-2}), and (3) the mandatory capacity to withstand attack (NIST SP 800-160 Vol. 2~\cite{nist800160v2}), supported by a fourth economic pillar: a federal resilience incentive structure designed to offset the operational costs of these rigorous constraints.

\section{Divergent Objectives: Safety Constraints vs. Security Controls}
\label{sec:divergence}

A fundamental safety gap exists in the built environment because current policy often conflates the data-centric goals of Information Technology (IT) with the physics-centric constraints of Operational Technology (OT). 
While used interchangeably in regulation, Cyber Security and Cyber Safety govern distinct system properties.
\textit{Security}, as defined in NIST SP 800-160 vol. 1~\cite{nist800160v1}, protects assets from adversarial compromise, prioritizing the CIA Triad (Confidentiality, Integrity, Availability).
In contrast, \textit{Safety} is an emergent property concerned with preventing loss of life, injury, or environmental damage~\cite{leveson2011}.

The conflict arises because standard IT security controls frequently enforce ``fail-secure'' states (e.g., locking access to protect data) that directly contradict the ``fail-safe'' requirements of physical systems (e.g., venting pressure or maintaining manual overrides).
As shown in Table~\ref{tab:conflicts}, compliant security controls like account lockouts and automated patching have actively introduced latent physical hazards into operational environments.
In these scenarios, the security control becomes the safety risk.

\begin{table}[htbp]
\caption{The Compliance Paradox: Security Controls vs. Safety Constraints}
\label{tab:conflicts}
\begin{center}
\begin{tabular}{|p{1.5cm}|p{1.5cm}|p{4.5cm}|}
\hline
\textbf{Security Control} & \textbf{Intended IT Benefit} & \textbf{Real-World Safety Failure} \\
\hline
\textbf{Account Lockout} \newline (3 failed attempts) & Prevents Brute Force Attacks & \textbf{KNX Devices (2023)}: Restrictive lockouts denied operator access to building control systems during emergency conditions \cite{CISA_KNX2023}. \\
\hline
\textbf{Fail-Secure Defaults} \newline (Electronic Locks) & Prevents Physical Theft & \textbf{Tesla (2025)}: Electronic door latches failed during battery fires, trapping occupants and delaying rescue (15 fatalities linked to egress failure) \cite{Bloomberg2025}. \\
\hline
\textbf{Automated Patching} & Vulnerability Remediation & \textbf{Advantech WebAccess}: Flawed patches introduced authentication bypasses and forced unplanned Supervisory Control and Data Acquisition (SCADA) downtime \cite{securityweek_advantech2023}. \\
\hline
\textbf{Encryption Overhead} & Data Confidentiality & \textbf{Real-Time Control}: Transport Layer Security (TLS) handshakes introduced latency exceeding the sub-100 ms budgets required by real-time safety loops~\cite{Hlayel2025}. \\
\hline
\end{tabular}
\end{center}
\end{table}

To resolve this, policy must shift the engineering objective from pure security towards \textit{Cyber Resilience}.
NIST SP 800-160 Vol. 2 defines resilience as the ability to ``anticipate, withstand, recover from, and adapt to'' adverse conditions~\cite{nist800160v2}. 
For the built environment, \textit{withstanding} is the critical differentiator. 
Unlike IT systems that can reboot to restore integrity, critical infrastructure must possess \textit{graceful extensibility}—the capacity to stretch under stress and continue operating in a degraded state to preserve life~\cite{Guttieri2025}.
This shift redefines the legal concept of ``Reasonable Care'' beyond simple checklist compliance to a duty of professional judgement, where engineers must demonstrate that they have actively identified and mitigated foreseeable cyber-physical risks~\cite{Niemeyer2024}. 

\section{Methodology}
\label{sec:method}

To determine how U.S. policy currently defines ``reasonable care'' for cyber-physical systems, we conducted a systematic mapping of critical infrastructure standards.

\subsection{Corpus Selection and Filtering}

\begin{figure}[htbp]
    \centering
    \includegraphics[width=\linewidth]{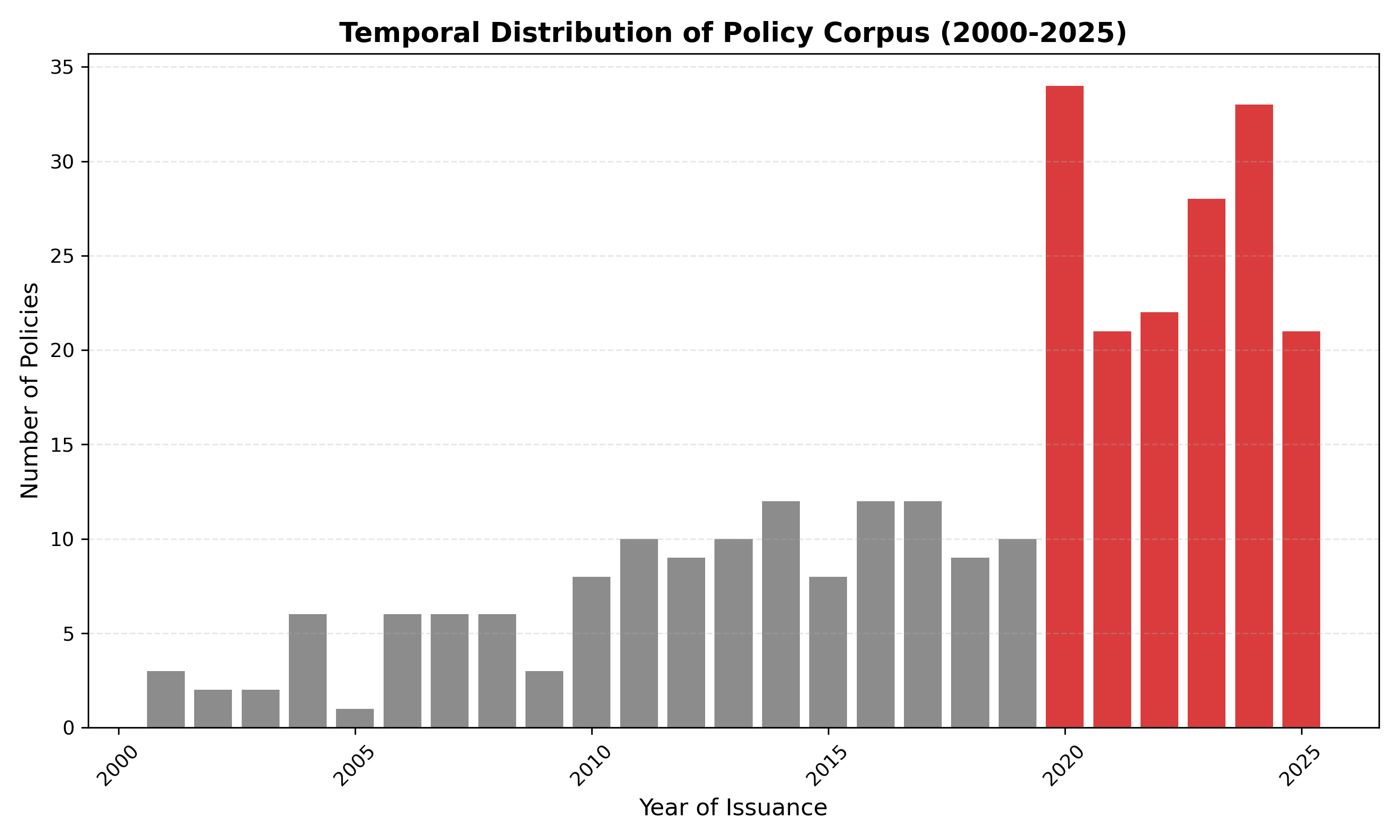}
    \caption{\textbf{Temporal Distribution of Policy Corpus.} The analysis highlights a significant acceleration in policy issuance from 2020–2025 (highlighted in red), reflecting increased federal focus on cyber-related policies.}
    \label{fig:year_dist}
\end{figure}

\begin{figure}[htbp]
    \centering
    \includegraphics[width=\linewidth]{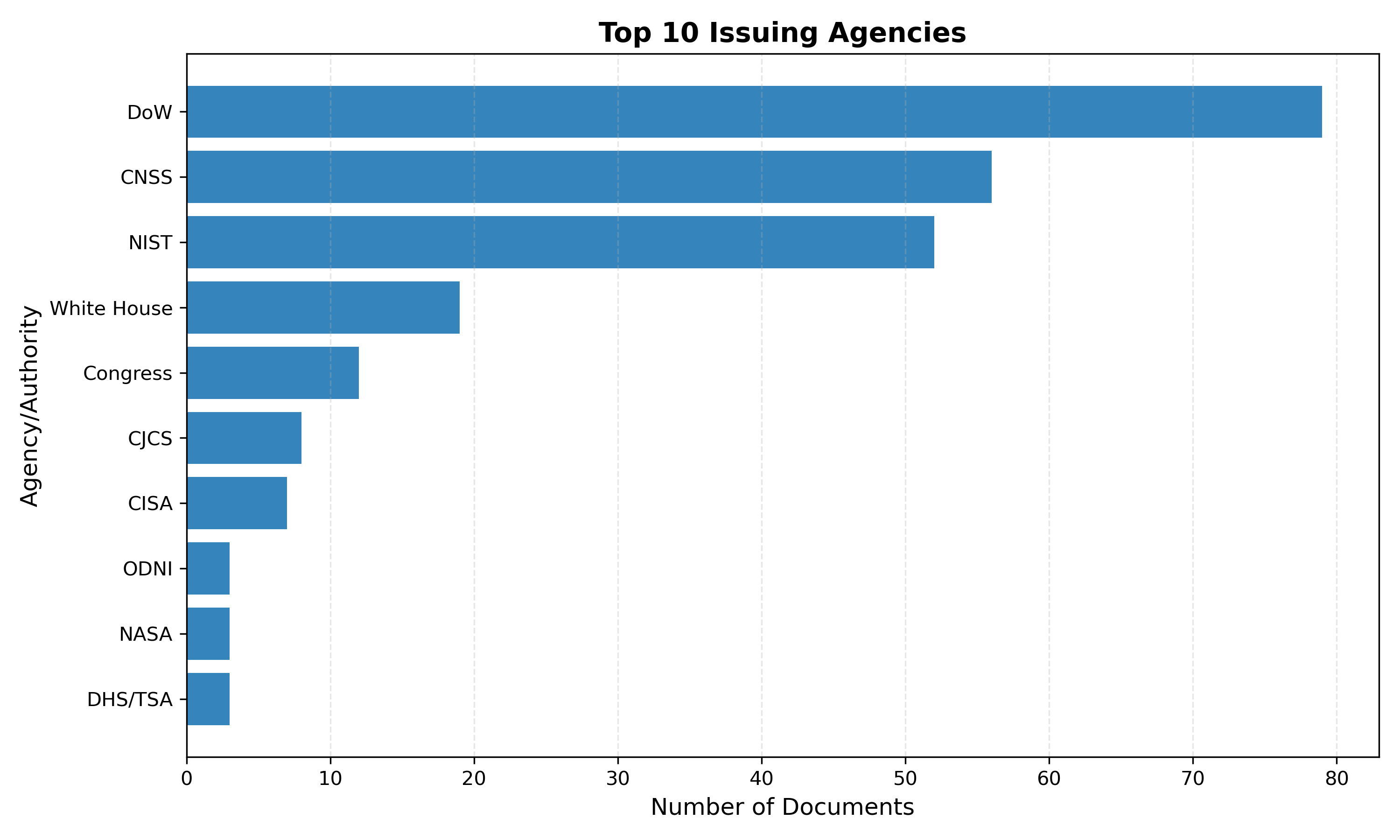}
    \caption{\textbf{Distribution by Issuing Authority.} The resulting corpus is heavily weighted toward defense and engineering authorities (DoW, CNSS, NIST).}
    \label{fig:agency_dist}
\end{figure}

We constructed the primary dataset ($N=292$) by extending the foundational, but single department, \textit{CSIAC DoW Cybersecurity Policy Chart} \cite{CSIAC_Policy_Chart} with operational guidance from the sixteen critical infrastructure sectors (e.g., \textit{America's Water Infrastructure Act}).
Because the CSIAC chart served as the seed corpus, the dataset skews toward Department of War (DoW) and CNSS issuances, with the remaining sectors represented by their principal sector-specific regulations.
We filtered this corpus using Boolean searches for keywords such as ``safety'' and ``resilience,'' verified applicability to the sixteen critical infrastructure sectors defined under PPD-21~\cite{ppd21} (a framework retained and superseded by NSM-22~\cite{NSM22} in 2024), and restricted the timeline to 2000--2025.
This process eliminated outdated guidance while capturing the significant acceleration in policy issuance illustrated in Figures \ref{fig:year_dist} and \ref{fig:agency_dist}.

\subsection{Systematic Policy Mapping}
\label{sec:Mapping}

We coded each document ($N=292$) across three analytical dimensions to evaluate alignment with Cyber Safety engineering requirements.

First, we coded the primary requirement of each document into six mutually exclusive categories to distinguish between static compliance activities and active systems engineering.
\textbf{Strategic Mandates} define the legal duty without the technical method (e.g., ``manage risk''), while \textbf{Administrative Compliance} focuses on bureaucratic artifacts such as workforce codes rather than system behavior.
\textbf{Lifecycle Management} covers process-oriented requirements tied to acquisition timelines (e.g., NIST RMF steps).
\textbf{Control Checklists} are enumerated lists of static security controls (e.g., NIST SP 800-53) focused on component configuration.
\textbf{Incident Coordination} mandates reactive reporting and command chains.
Finally, \textbf{Engineering Constraints} are technical requirements linking controls to physics-based hazards or firm architectural constraints (e.g., Anti-Tamper, TEMPEST, FDA Medical Device guidance).

Second, we mapped the operational intent of each document to the four phases of the NIST SP 800-160 Vol. 2 resilience lifecycle: \textbf{Anticipate} (threat modeling and planning), \textbf{Withstand} (maintaining essential functions during attack), \textbf{Recover} (restoration of services), and \textbf{Adapt} (forensic redesign).
Simultaneously, we classified the target environment to isolate the distinct requirements of the built environment.
Policies were coded as \textbf{Enterprise IT} if they governed general-purpose business systems (confidentiality-focused) or \textbf{Cyber-Physical Systems (CPS)} if they governed systems where real-time availability and safety are the primary criteria (e.g., ICS, weapons systems).
This mapping reveals the structural emphasis of the current regulatory environment for both IT and CPS (see Figure \ref{fig:heatmap_full}).

\begin{figure}[htbp]
\centering
\begin{tikzpicture}[font=\sffamily\footnotesize, scale=0.55]

\def\cellwidth{2.8} 
\def\cellheight{1.2} 

\node[align=center, anchor=south, font=\bfseries] at (0.5*\cellwidth, 6*\cellheight) {\footnotesize Anticipate};
\node[align=center, anchor=south, font=\bfseries] at (1.5*\cellwidth, 6*\cellheight) {\footnotesize Withstand};
\node[align=center, anchor=south, font=\bfseries] at (2.5*\cellwidth, 6*\cellheight) {\footnotesize Recover};
\node[align=center, anchor=south, font=\bfseries] at (3.5*\cellwidth, 6*\cellheight) {\footnotesize Adapt};

\node[align=right, anchor=east, font=\bfseries] at (0, 5.5*\cellheight) {\footnotesize Admin Compl.};
\node[align=right, anchor=east, font=\bfseries] at (0, 4.5*\cellheight) {\footnotesize Lifecycle Mgmt.};
\node[align=right, anchor=east, font=\bfseries] at (0, 3.5*\cellheight) {\footnotesize Incident Coord.};
\node[align=right, anchor=east, font=\bfseries] at (0, 2.5*\cellheight) {\footnotesize Control Checklist};
\node[align=right, anchor=east, font=\bfseries] at (0, 1.5*\cellheight) {\footnotesize Eng. Constraint};


\fill[red!70] (0, 5*\cellheight) rectangle (\cellwidth, 6*\cellheight); 
\draw[thick] (0, 5*\cellheight) rectangle (\cellwidth, 6*\cellheight); 
\node[align=center] at (0.5*\cellwidth, 5.5*\cellheight) {$n=48$\\ \scriptsize\textit{NSM-22}};

\fill[yellow!20] (\cellwidth, 5*\cellheight) rectangle (2*\cellwidth, 6*\cellheight); 
\draw[thick] (\cellwidth, 5*\cellheight) rectangle (2*\cellwidth, 6*\cellheight); 
\node[align=center] at (1.5*\cellwidth, 5.5*\cellheight) {$n=4$\\ \scriptsize\textit{TSA SD-02C}};

\fill[yellow!10] (2*\cellwidth, 5*\cellheight) rectangle (3*\cellwidth, 6*\cellheight); 
\draw[thick] (2*\cellwidth, 5*\cellheight) rectangle (3*\cellwidth, 6*\cellheight); 
\node[align=center] at (2.5*\cellwidth, 5.5*\cellheight) {$n=1$\\ \scriptsize\textit{DoDD 3020}};

\fill[white] (3*\cellwidth, 5*\cellheight) rectangle (4*\cellwidth, 6*\cellheight); 
\draw[thick] (3*\cellwidth, 5*\cellheight) rectangle (4*\cellwidth, 6*\cellheight); 
\node[align=center] at (3.5*\cellwidth, 5.5*\cellheight) {$n=0$};

\fill[red!60] (0, 4*\cellheight) rectangle (\cellwidth, 5*\cellheight); 
\draw[thick] (0, 4*\cellheight) rectangle (\cellwidth, 5*\cellheight); 
\node[align=center] at (0.5*\cellwidth, 4.5*\cellheight) {$n=40$\\ \scriptsize\textit{AWIA 2018}};

\fill[yellow!30] (\cellwidth, 4*\cellheight) rectangle (2*\cellwidth, 5*\cellheight); 
\draw[thick] (\cellwidth, 4*\cellheight) rectangle (2*\cellwidth, 5*\cellheight); 
\node[align=center] at (1.5*\cellwidth, 4.5*\cellheight) {$n=5$\\ \scriptsize\textit{CNSSD 504}};

\fill[white] (2*\cellwidth, 4*\cellheight) rectangle (3*\cellwidth, 5*\cellheight); 
\draw[thick] (2*\cellwidth, 4*\cellheight) rectangle (3*\cellwidth, 5*\cellheight); 
\node[align=center] at (2.5*\cellwidth, 4.5*\cellheight) {$n=0$};

\fill[yellow!20] (3*\cellwidth, 4*\cellheight) rectangle (4*\cellwidth, 5*\cellheight); 
\draw[thick] (3*\cellwidth, 4*\cellheight) rectangle (4*\cellwidth, 5*\cellheight); 
\node[align=center] at (3.5*\cellwidth, 4.5*\cellheight) {$n=3$\\ \scriptsize\textit{DoDI 5000.02}};

\fill[orange!20] (0, 3*\cellheight) rectangle (\cellwidth, 4*\cellheight); 
\draw[thick] (0, 3*\cellheight) rectangle (\cellwidth, 4*\cellheight); 
\node[align=center] at (0.5*\cellwidth, 3.5*\cellheight) {$n=11$\\ \scriptsize\textit{DHS STCP}};

\fill[white] (\cellwidth, 3*\cellheight) rectangle (2*\cellwidth, 4*\cellheight); 
\draw[thick] (\cellwidth, 3*\cellheight) rectangle (2*\cellwidth, 4*\cellheight); 
\node[align=center] at (1.5*\cellwidth, 3.5*\cellheight) {$n=0$};

\fill[orange!20] (2*\cellwidth, 3*\cellheight) rectangle (3*\cellwidth, 4*\cellheight); 
\draw[thick] (2*\cellwidth, 3*\cellheight) rectangle (3*\cellwidth, 4*\cellheight); 
\node[align=center] at (2.5*\cellwidth, 3.5*\cellheight) {$n=10$\\ \scriptsize\textit{CIRCIA}};

\fill[yellow!20] (3*\cellwidth, 3*\cellheight) rectangle (4*\cellwidth, 4*\cellheight); 
\draw[thick] (3*\cellwidth, 3*\cellheight) rectangle (4*\cellwidth, 4*\cellheight); 
\node[align=center] at (3.5*\cellwidth, 3.5*\cellheight) {$n=2$\\ \scriptsize\textit{FISMA 2014}};

\fill[orange!30] (0, 2*\cellheight) rectangle (\cellwidth, 3*\cellheight); 
\draw[thick] (0, 2*\cellheight) rectangle (\cellwidth, 3*\cellheight); 
\node[align=center] at (0.5*\cellwidth, 2.5*\cellheight) {$n=17$\\ \scriptsize\textit{CISA CPG}};

\fill[red!50] (\cellwidth, 2*\cellheight) rectangle (2*\cellwidth, 3*\cellheight); 
\draw[thick] (\cellwidth, 2*\cellheight) rectangle (2*\cellwidth, 3*\cellheight); 
\node[align=center] at (1.5*\cellwidth, 2.5*\cellheight) {$n=39$\\ \scriptsize\textit{CMMC}};

\fill[white] (2*\cellwidth, 2*\cellheight) rectangle (3*\cellwidth, 3*\cellheight); 
\draw[thick] (2*\cellwidth, 2*\cellheight) rectangle (3*\cellwidth, 3*\cellheight); 
\node[align=center] at (2.5*\cellwidth, 2.5*\cellheight) {$n=0$};

\fill[white] (3*\cellwidth, 2*\cellheight) rectangle (4*\cellwidth, 3*\cellheight); 
\draw[thick] (3*\cellwidth, 2*\cellheight) rectangle (4*\cellwidth, 3*\cellheight); 
\node[align=center] at (3.5*\cellwidth, 2.5*\cellheight) {$n=0$};

\fill[red!30] (0, 1*\cellheight) rectangle (\cellwidth, 2*\cellheight); 
\draw[thick] (0, 1*\cellheight) rectangle (\cellwidth, 2*\cellheight); 
\node[align=center] at (0.5*\cellwidth, 1.5*\cellheight) {$n=23$\\ \scriptsize\textit{CISA SbD}};

\fill[orange!20] (\cellwidth, 1*\cellheight) rectangle (2*\cellwidth, 2*\cellheight); 
\draw[thick] (\cellwidth, 1*\cellheight) rectangle (2*\cellwidth, 2*\cellheight); 
\node[align=center] at (1.5*\cellwidth, 1.5*\cellheight) {$n=12$\\ \scriptsize\textit{10CFR73.54}};

\fill[yellow!10] (2*\cellwidth, 1*\cellheight) rectangle (3*\cellwidth, 2*\cellheight); 
\draw[thick] (2*\cellwidth, 1*\cellheight) rectangle (3*\cellwidth, 2*\cellheight); 
\node[align=center] at (2.5*\cellwidth, 1.5*\cellheight) {$n=1$\\ \scriptsize\textit{CNSSI-1015}};

\fill[yellow!20] (3*\cellwidth, 1*\cellheight) rectangle (4*\cellwidth, 2*\cellheight); 
\draw[thick] (3*\cellwidth, 1*\cellheight) rectangle (4*\cellwidth, 2*\cellheight); 
\node[align=center] at (3.5*\cellwidth, 1.5*\cellheight) {$n=2$\\ \tiny\textit{FERC 18CFR12}};
\end{tikzpicture}
\caption{\textbf{Distribution of the Full Policy Corpus (All Domains).} This heat map visualizes the frequency of mandates ($n=218$) across the NIST resilience lifecycle for the entire filtered dataset, excluding high-level strategic mandates. The distribution reveals a strong bias toward the \textit{Anticipate} phase ($n=139$), dominated by administrative compliance and lifecycle management obligations. This represents the broad cyber regulatory volume, including generic enterprise IT governance.}
\label{fig:heatmap_full}
\end{figure}

Finally, to resolve the ambiguity of regulations that act as administrative ``shells,'' we performed a \textit{Pointer-Target} analysis. We identified documents acting as ``Pointers'' (e.g., \textit{10 CFR 73.54}) that mandate external ``Targets'' (e.g., \textit{NIST SP 800-53}). We coded obligations based on the engineering rigor of the target standard to ensure the analysis reflects the technical reality rather than the administrative language.

\subsection{Analytical Constraints}
The corpus was constructed and coded by the authors. 
This methodology maps regulatory text by mention frequency, not organizational implementation or adoption maturity. 
Policies drafted for IT environments are coded as written, without assuming CPS-specific operational adjustments, and the scope is restricted to the U.S. federal corpus. 
A sector-balanced expansion, including release of the full coded dataset, is reserved for a forthcoming extended study.

\section{Findings and Proposed Standard of Care}
\label{sec:findings_and_care}

The application of the policy mapping framework reveals a critical divergence: while the volume of cybersecurity guidance has exploded, the core definition of ``reasonable care'' remains misaligned with verifiable safety outcomes.
As shown in Figure \ref{fig:heatmap_cps_both}, federal policy has successfully universalized \textit{Administrative Compliance} and \textit{Incident Coordination}, effectively defining care as the ability to document defenses and report failures.
However, this regime emphasizes governance artifacts at the expense of engineering evidence about system behavior.
To resolve this, we identify three major disconnects and propose a modernized standard of care that re-anchors liability protection in verifiable engineering artifacts.

\begin{figure}[htbp]
\centering
\begin{tikzpicture}[font=\sffamily\footnotesize, scale=0.55]

\def\cellwidth{2.8} 
\def\cellheight{1.2} 

\node[align=center, anchor=south, font=\bfseries] at (0.5*\cellwidth, 6*\cellheight) {\footnotesize Anticipate};
\node[align=center, anchor=south, font=\bfseries] at (1.5*\cellwidth, 6*\cellheight) {\footnotesize Withstand};
\node[align=center, anchor=south, font=\bfseries] at (2.5*\cellwidth, 6*\cellheight) {\footnotesize Recover};
\node[align=center, anchor=south, font=\bfseries] at (3.5*\cellwidth, 6*\cellheight) {\footnotesize Adapt};

\node[align=right, anchor=east, font=\bfseries] at (0, 5.5*\cellheight) {\footnotesize Admin Compl.};
\node[align=right, anchor=east, font=\bfseries] at (0, 4.5*\cellheight) {\footnotesize Lifecycle Mgmt.};
\node[align=right, anchor=east, font=\bfseries] at (0, 3.5*\cellheight) {\footnotesize Incident Coord.};
\node[align=right, anchor=east, font=\bfseries] at (0, 2.5*\cellheight) {\footnotesize Control Checklist};
\node[align=right, anchor=east, font=\bfseries] at (0, 1.5*\cellheight) {\footnotesize Eng. Constraint};


\fill[red!30] (0, 5*\cellheight) rectangle (\cellwidth, 6*\cellheight);
\draw[thick] (0, 5*\cellheight) rectangle (\cellwidth, 6*\cellheight);
\node[align=center] at (0.5*\cellwidth, 5.5*\cellheight) {$n=23$\\ \scriptsize\textit{NSM-22}};

\fill[yellow!20] (\cellwidth, 5*\cellheight) rectangle (2*\cellwidth, 6*\cellheight);
\draw[thick] (\cellwidth, 5*\cellheight) rectangle (2*\cellwidth, 6*\cellheight);
\node[align=center] at (1.5*\cellwidth, 5.5*\cellheight) {$n=3$\\ \scriptsize\textit{TSA SD-02C}};

\fill[white] (2*\cellwidth, 5*\cellheight) rectangle (3*\cellwidth, 6*\cellheight);
\draw[thick] (2*\cellwidth, 5*\cellheight) rectangle (3*\cellwidth, 6*\cellheight);
\node[align=center] at (2.5*\cellwidth, 5.5*\cellheight) {$n=0$};

\fill[white] (3*\cellwidth, 5*\cellheight) rectangle (4*\cellwidth, 6*\cellheight);
\draw[thick] (3*\cellwidth, 5*\cellheight) rectangle (4*\cellwidth, 6*\cellheight);
\node[align=center] at (3.5*\cellwidth, 5.5*\cellheight) {$n=0$};

\fill[orange!20] (0, 4*\cellheight) rectangle (\cellwidth, 5*\cellheight);
\draw[thick] (0, 4*\cellheight) rectangle (\cellwidth, 5*\cellheight);
\node[align=center] at (0.5*\cellwidth, 4.5*\cellheight) {$n=12$\\ \scriptsize\textit{AWIA 2018}};

\fill[yellow!20] (\cellwidth, 4*\cellheight) rectangle (2*\cellwidth, 5*\cellheight);
\draw[thick] (\cellwidth, 4*\cellheight) rectangle (2*\cellwidth, 5*\cellheight);
\node[align=center] at (1.5*\cellwidth, 4.5*\cellheight) {$n=2$\\ \scriptsize\textit{CNSSD 504}};

\fill[white] (2*\cellwidth, 4*\cellheight) rectangle (3*\cellwidth, 5*\cellheight); 
\draw[thick] (2*\cellwidth, 4*\cellheight) rectangle (3*\cellwidth, 5*\cellheight);
\node[align=center] at (2.5*\cellwidth, 4.5*\cellheight) {$n=0$};

\fill[yellow!10] (3*\cellwidth, 4*\cellheight) rectangle (4*\cellwidth, 5*\cellheight);
\draw[thick] (3*\cellwidth, 4*\cellheight) rectangle (4*\cellwidth, 5*\cellheight);
\node[align=center] at (3.5*\cellwidth, 4.5*\cellheight) {$n=1$\\ \tiny\textit{PHMSA API-1173}};

\fill[yellow!30] (0, 3*\cellheight) rectangle (\cellwidth, 4*\cellheight);
\draw[thick] (0, 3*\cellheight) rectangle (\cellwidth, 4*\cellheight);
\node[align=center] at (0.5*\cellwidth, 3.5*\cellheight) {$n=6$\\ \scriptsize\textit{DHS STCP}};

\fill[white] (\cellwidth, 3*\cellheight) rectangle (2*\cellwidth, 4*\cellheight);
\draw[thick] (\cellwidth, 3*\cellheight) rectangle (2*\cellwidth, 4*\cellheight);
\node[align=center] at (1.5*\cellwidth, 3.5*\cellheight) {$n=0$};

\fill[yellow!30] (2*\cellwidth, 3*\cellheight) rectangle (3*\cellwidth, 4*\cellheight);
\draw[thick] (2*\cellwidth, 3*\cellheight) rectangle (3*\cellwidth, 4*\cellheight);
\node[align=center] at (2.5*\cellwidth, 3.5*\cellheight) {$n=7$\\ \scriptsize\textit{CIRCIA}};

\fill[white] (3*\cellwidth, 3*\cellheight) rectangle (4*\cellwidth, 4*\cellheight);
\draw[thick] (3*\cellwidth, 3*\cellheight) rectangle (4*\cellwidth, 4*\cellheight);
\node[align=center] at (3.5*\cellwidth, 3.5*\cellheight) {$n=0$};

\fill[orange!20] (0, 2*\cellheight) rectangle (\cellwidth, 3*\cellheight);
\draw[thick] (0, 2*\cellheight) rectangle (\cellwidth, 3*\cellheight);
\node[align=center] at (0.5*\cellwidth, 2.5*\cellheight) {$n=10$\\ \scriptsize\textit{CISA CPG}};

\fill[red!40] (\cellwidth, 2*\cellheight) rectangle (2*\cellwidth, 3*\cellheight);
\draw[thick] (\cellwidth, 2*\cellheight) rectangle (2*\cellwidth, 3*\cellheight);
\node[align=center] at (1.5*\cellwidth, 2.5*\cellheight) {$n=29$\\ \scriptsize\textit{CMMC}};

\fill[white] (2*\cellwidth, 2*\cellheight) rectangle (3*\cellwidth, 3*\cellheight);
\draw[thick] (2*\cellwidth, 2*\cellheight) rectangle (3*\cellwidth, 3*\cellheight);
\node[align=center] at (2.5*\cellwidth, 2.5*\cellheight) {$n=0$};

\fill[white] (3*\cellwidth, 2*\cellheight) rectangle (4*\cellwidth, 3*\cellheight);
\draw[thick] (3*\cellwidth, 2*\cellheight) rectangle (4*\cellwidth, 3*\cellheight);
\node[align=center] at (3.5*\cellwidth, 2.5*\cellheight) {$n=0$};

\fill[orange!30] (0, 1*\cellheight) rectangle (\cellwidth, 2*\cellheight);
\draw[thick] (0, 1*\cellheight) rectangle (\cellwidth, 2*\cellheight);
\node[align=center] at (0.5*\cellwidth, 1.5*\cellheight) {$n=17$\\ \tiny\textit{FAA AC-20 115D}};

\fill[orange!20] (\cellwidth, 1*\cellheight) rectangle (2*\cellwidth, 2*\cellheight);
\draw[thick] (\cellwidth, 1*\cellheight) rectangle (2*\cellwidth, 2*\cellheight);
\node[align=center] at (1.5*\cellwidth, 1.5*\cellheight) {$n=12$\\ \scriptsize\textit{10CFR73.54}};

\fill[yellow!10] (2*\cellwidth, 1*\cellheight) rectangle (3*\cellwidth, 2*\cellheight);
\draw[thick] (2*\cellwidth, 1*\cellheight) rectangle (3*\cellwidth, 2*\cellheight);
\node[align=center] at (2.5*\cellwidth, 1.5*\cellheight) {$n=1$\\ \scriptsize\textit{CNSSI-1015}};

\fill[yellow!20] (3*\cellwidth, 1*\cellheight) rectangle (4*\cellwidth, 2*\cellheight);
\draw[thick] (3*\cellwidth, 1*\cellheight) rectangle (4*\cellwidth, 2*\cellheight);
\node[align=center] at (3.5*\cellwidth, 1.5*\cellheight) {$n=2$\\ \tiny\textit{FERC 18CFR12}};
\end{tikzpicture}
\caption{\textbf{Distribution of Policy Clauses by Requirement Type for CPS in critical infrastructure (the built environment).} This heat map visualizes the frequency of clauses ($n=125$) across five requirement categories (rows) and four resilience phases (columns). Darker cells indicate a higher density of mandates.}
\label{fig:heatmap_cps_both}
\end{figure}

\subsection{Finding 1: The ``Delegated'' Defense Strategy}

The dataset reveals a distinct ``Delegated'' defense posture in the \textit{Withstand} phase. As shown in Table \ref{tab:dual_payload}, while the majority of policy mandates are concentrated in the \textit{Anticipate} phase (dominated by administrative preparation), the mission-critical \textit{Withstand} phase ($n=46$) shifts strategy. 
Of these \textit{Withstand} obligations, 87\% utilize ``pointers'' that link legal compliance to external technical standards rather than defining specific performance criteria.

This delegation strategy is not inherently flawed, it leverages the specialized expertise of standard-setting bodies. 
However, analysis reveals a critical calibration failure in the selection of these targets:
\begin{itemize}
    \item \textbf{The Miscalibrated Pointer (61\%):} The majority of pointers ($n=28$) delegate defense to generic IT frameworks (e.g., NIST SP 800-53). These references are effectively hollow for CPS: they enforce data confidentiality rules (e.g., logging, passwords) on systems that require physical resiliency. While excellent for IT data, they lack the hazard analysis requirements necessary for the built environment.
    \item \textbf{The Resilient Pointer (26\%):} A minority ($n=12$) delegates to domain-specific engineering standards (e.g., International Society of Automation (ISA) / International Electrotechnical Commission (IEC) 62443~\cite{ISA62443}, IMO MSC.428~\cite{IMO_MSC428}). We note that these standards do prescribe the missing elements of reasonable care. For example, ISA/IEC 62443-3-2 mandates risk assessment based on physical segmentation (Zones/Conduits), and IMO MSC.428 ties cyber risk directly to the vessel's Safety Management System (SMS).
\end{itemize}

\begin{table}[htbp]
\caption{Delegated Defense Standards and Their Treatment of ``Reasonable Care''}
\begin{center}
\begin{tabular}{|p{0.20\linewidth}|p{0.18\linewidth}|p{0.45\linewidth}|}
\hline
&\multicolumn{2}{|c|}{\textbf{Delegated Standard Details}} \\
\cline{2-3} 
\textbf{Obligation Type} & \textbf{\textit{Primary Targets}}& \textbf{\textit{Nature of ``Reasonable Care''}} \\
\hline
\textbf{Control Checklists} & \textbf{NIST SP 800-53} \newline (IT General) \newline \textbf{NIST SP 800-171} \newline (CUI Data) & \textbf{Administrative}: Mandates audit trails, password complexity, and boundary protection. \newline \textit{Gap:} Prescribes controls without requiring a hazard analysis to justify them. \\
\hline
\textbf{Engineering Constraints} & \textbf{ISA/IEC 62443} \newline (Industrial) \newline \textbf{IMO MSC.428} \newline (Maritime) & \textbf{Functional}: Mandates safe-state failures, zone partitioning, and determinable degradation. \newline \textit{Value:} Explicitly links security controls to physical consequence (e.g., HAZOP/SMS). \\
\hline
\end{tabular}
\label{tab:dual_payload}
\end{center}
\end{table}

\vspace{0.5em} \noindent
\textbf{Proposed Standard I: Hazard-Specific Traceability} 
A valid standard of care must require hazard-specific traceability. 
Policy should mandate the risk assessment framework of engineering standards (such as \textbf{ISA/IEC 62443-3-2}~\cite{ISA62443}) that require partitioning systems based on physical consequence rather than network connectivity.
However, because traditional component-based assessments cannot model complex system interactions, regulation must endorse a holistic, systems-based methodology, such as \textbf{Cyber-Informed Engineering (CIE)~\cite{Bochman2021}} or \textbf{System-Theoretic Process Analysis (STPA)~\cite{leveson2018stpa,Young2013}}, to derive the underlying hazard scenarios.
Reasonable care is demonstrated only when an asset owner can prove, via bi-directional traceability, that a digital control was selected explicitly to mitigate/prevent a physical consequence identified in a \textbf{Consequence-Driven Cyber-Informed Engineering (CCE)} or systems-theoretic analysis.
\vspace{0.5em}

\subsection{Finding 2: The ``Misplaced'' Recovery Standard}

Current policy creates a resilience gap in the \textit{Recover} phase ($n=8$ for CPS) by substituting ``Notification'' for ``Navigation.''
Our analysis shows that CPS \textit{Recover} obligations consist almost exclusively ($88\%$, $n=7$) of \textit{Incident Coordination} mandates (e.g., \textit{PPD-41, Cyber Incident Reporting for Critical Infrastructure Act (CIRCIA)}).
These policies enforce a standard of informing the government of failure (Notification) or executing organizational Continuity of Operations Plans (COOP), but fail to enforce a standard for engineering the system to safely override or degrade during that failure.
While COOP mandates ensure personnel know where to go during a crisis, they rarely prescribe the technical logic required for a cyber-physical system to recover its safe state without digital connectivity.
The single \textit{Engineering Constraint} in this phase ($n=1$) shows where the line is currently drawn.
The Committee on National Security Systems Instruction (CNSSI-1015) does its own job well, standardizing the generation and retention of audit logs so that an incident can be reconstructed after the fact.
That evidence is necessary, but it sits alongside the recovery problem rather than solving it.
Nothing in the \textit{Recover} phase obligates an operator to engineer the system so that it can reach a safe state on its own when the digital layer is degraded or gone.

The engineering that real recovery would require, such as fail-safe logic and determinable degradation, does exist in the corpus, but it lives in a narrow place.
It appears inside the small set of delegated technical standards identified in Finding 1, where references like ISA/IEC 62443~\cite{ISA62443} and IMO MSC.428~\cite{IMO_MSC428} do call for safe-state behavior and graceful degradation.
The gap is one of reach, not knowledge.
Those standards bind only the operators who fall under a regulator that points to them, while the generally applicable Recover obligations across the corpus ask for notification and continuity planning rather than for evidence that the system itself can return to a safe state.
For most asset owners, then, engineered recovery is available in principle but never actually required by policy.

By treating \textit{Recover} as a reporting step, the broadly binding policy leaves the connection between an incident and the engineering work done to survive it unmade, so an operator can satisfy its recovery duties without ever showing the system was built to ride through the event.
We note a methodological caveat.
Because each clause is mapped to a single resilience phase, some safe-state requirements that a reader might place in \textit{Recover} were coded under \textit{Withstand}.
This boundary is a feature of the coding scheme rather than of the standards, which treat withstanding and recovering as a continuum.
The finding does not rest on that boundary.
It rests on the observation that the broadly binding obligations are notification while the engineered safe-state requirements are confined to a small set of delegated standards.

\vspace{0.5em} \noindent
\textbf{Proposed Standard II: The Assurance Case as the Recovery Link} 
To close the verification gap, regulation must mandate \textit{structured assurance cases} (e.g., \textbf{ISO/IEC/IEEE 15026-2}) that link \textit{Recover} obligations back to \textit{Withstand} engineering constraints.
An assurance case acts as the ``Navigation Plan'' for recovery, allowing an operator to claim reasonable care not by showing a phone log to the Cybersecurity and Infrastructure Security Agency (CISA) (Notification), but by organizing existing engineering evidence into a cohesive argument proving the system is designed to recover to a safe state.
\vspace{0.5em}

\subsection{Finding 3: The Adaptation Anomaly}
The final phase of the resilience lifecycle, \textit{Adapt} ($n=3$), reveals a critical divergence in how ``Reasonable Care'' is applied. This phase contains the only significant ``Right-of-Boom'' engineering constraints ($n=2$ for the built environment), and they are exclusively driven by established sector-specific regulators: TSA SD 1582-21-01~\cite{TSA_SD_1582_21_01} for public transportation and passenger rail and FERC 18 CFR Part 12 for dams.

Our data highlights a possible regression in the rail industry: the 2024 Transportation Security Administration (TSA) Surface Cyber Risk Management (SCRM) Notice of Proposed Rulemaking (NPRM)~\cite{TSA_NPRM_2024} falls into the \textit{Anticipate/Control Checklist} category.
While the preceding SD 1582-21-01 explicitly mandated network segmentation, the NPRM replaces this hard constraint with a ``performance-based'' Cyber Risk Management Program~\cite{GrandTrunkvTSA2024}. By allowing operators to substitute physical adaptation architectures with administrative remediation plans (Corrective Action Plans), the NPRM effectively sunsets the requirement for resilience.

This confirms that ``Reasonable Care'' is currently a function of regulatory geography. 
If an asset falls under the jurisdiction of a safety-critical regulator such as the Federal Energy Regulatory Commission (FERC), it is required to be resilient. 
If it falls under general cyber policy, where the permanent standard is merely a plan, then it is only required to be compliant.

\vspace{0.5em} \noindent
\textbf{Proposed Standard III: Cyber Resiliency Engineering} 
Reasonable care must mandate the capacity to \textit{withstand} and \textit{adapt} to attacks: reporting incidents is necessary, but regulations should also require that systems maintain or return to a safe state during and after an incident.
We propose that the standard of care for critical infrastructure must align with \textbf{NIST SP 800-160 Vol. 2 (Cyber Resiliency Engineering)}. 
This requires engineering ``Left-of-Boom'' capabilities (e.g., adaptive response) and explicitly mandating \textit{non-digital fallbacks} (e.g., mechanical interlocks and analog governors) that provide deterministic safety guarantees, ensuring that the system can ``fight through'' a compromise without entering a hazardous state.
\vspace{0.5em} 

\subsection{Further Recommendation: Incentivizing the Cost of Care}
\label{sec:incentives}

Implementing the engineering-based standard of care proposed in the findings imposes a ``safety premium.''
Because resilient architectures (e.g., segmentation, diversity) are inherently less efficient than converged IT networks, the market will naturally optimize for fragility.
To ensure the safe state is a viable business decision, federal policy must bridge this cost delta by incentivizing the specific technologies that enable systems to \textit{withstand} and \textit{adapt} to compromise. 
A ``Resilience Incentive Program'' should subsidize the high-friction architectures that compliance checklists ignore, specifically deterministic out-of-band recovery networks, hardware-enforced segmentation (e.g., unidirectional gateways), and non-digital fallbacks (e.g., analog governors).
By offsetting the lifecycle costs of these right-of-boom capabilities, the government can make it financially feasible for operators to deploy resilient architectures.

\vspace{0.5em} \noindent
\textbf{Policy Takeaway:} A higher standard of care requires a lower barrier to entry. 
Policymakers must incentivize technologies that reduce the friction of \textit{Withstand} and \textit{Adapt} capabilities, ensuring that the ``Safe State'' becomes not just a legal obligation, but a viable business decision.
\vspace{0.5em}

\section{Conclusion}
\label{sec:conclusion}

Current U.S. cyber policy, centered on security, gives a false impression that safety in cyber-physical systems is assured.
By prioritizing administrative compliance, the federal government has established a regime where critical infrastructure owners can demonstrate formal diligence while their systems still lack the ability to withstand disruption. 
Our analysis confirms that the current ``Standard of Care'' is defined as the ability to document defenses and report failures, rather than engineering to survive them.

For the built environment, this must change. 
As the kinetic consequences of cyber insecurity escalate, reasonable care must evolve from a static checklist to a dynamic engineering discipline. 
We must move beyond validating the \textit{presence} of digital controls (e.g., ``is the firewall installed?'') to verifying the \textit{preservation} of safety properties (e.g., ``will the system fail safe if the firewall is breached?'').
This demands a standard anchored in three engineering imperatives: \textbf{Traceability} from hazards to requirements, \textbf{Structured Assurance} cases that evolve with the threat, and \textbf{Resiliency} that guarantees physical safety through non-digital fallbacks.

Implementing this engineering rigor imposes a ``safety premium,'' and federal policy must incentivize the necessary investment.
But ultimately, we cannot paper over the risks of the physical world. 
If the United States intends to secure its critical infrastructure against determined adversaries, it must align regulatory obligations with the physical behavior of systems rather than solely with documentation practices.

\bibliographystyle{IEEEtran}
\bibliography{refs} 

\appendices
\section{Policy Acronyms}
\label{sec:acronyms}
The following acronyms denote the policy instruments referenced in the figures and findings.

\begin{tabular}{p{0.25\columnwidth}p{0.71\columnwidth}}
\textbf{10 CFR 73.54} & Protection of Digital Computer and Communication Systems (Nuclear) \\
\textbf{AWIA} & America's Water Infrastructure Act (Sec.\ 2013) \\
\textbf{CIRCIA} & Cyber Incident Reporting for Critical Infrastructure Act \\
\textbf{CISA CPG} & Cross-Sector Cybersecurity Performance Goals \\
\textbf{CISA SbD} & CISA Secure by Design Pledge \\
\textbf{CMMC} & Cybersecurity Maturity Model Certification \\
\textbf{CNSSD 504} & Protecting National Security Systems from Insider Threat \\
\textbf{CNSSI-1015} & Enterprise Audit Management for National Security Systems \\
\textbf{DHS STCP} & DHS Soft Targets and Crowded Places Resources \\
\textbf{DoDD 3020.40} & Mission Assurance \\
\textbf{DoDI 5000.02} & Operation of the Adaptive Acquisition Framework \\
\textbf{FAA AC-20 115D} & Airborne Software Development Assurance \\
\textbf{FERC 18 CFR Pt.\ 12} & Safety of Water Power Projects (Dam Safety) \\
\textbf{FISMA} & Federal Information Security Modernization Act of 2014 \\
\textbf{NSM-22} & National Security Memorandum on Critical Infrastructure Security and Resilience \\
\textbf{PHMSA API 1173} & Pipeline Safety Management Systems (API RP 1173) \\
\textbf{PPD-21} & Critical Infrastructure Security and Resilience (superseded by NSM-22) \\
\textbf{PPD-41} & United States Cyber Incident Coordination \\
\textbf{TSA SD-02C} & TSA Security Directive Pipeline-2021-02C \\
\textbf{TSA SD 1582-21-01} & TSA Enhancing Public Transportation and Passenger Railroad Cybersecurity \\
\end{tabular}

\end{document}